\documentclass[prd,twocolumn,showpacs]{revtex4}
\usepackage[english]{babel}
\usepackage{amsmath}
\usepackage{amsfonts}
\usepackage{amssymb}
\usepackage{mathrsfs}
\usepackage{txfonts}
\usepackage{graphicx}
\usepackage{dcolumn}
\usepackage{bm}

\begin{document}

\title{The effective dynamics of loop quantum $R^2$ cosmology}

\author{Long Chen}
\email{chen_long@mail.bnu.edu.cn}
\affiliation{College of Physics and Electrical Engineering, Xinyang Normal
  University, Xinyang, 464000, Henan, China}
\date{\today}

\begin{abstract}
The effective dynamics of loop quantum $f (R)$ cosmology in Jordan frame is
considered by using the dynamical system method and numerical method. To make the analyze
in detail, we focus on $R^2$ model since it is simple and favored from observations.
In classical theory, $(\phi = 1, \dot{\phi} = 0)$ is the unique fixed point in both contracting  and expanding states, and all solutions are either starting from the fixed point or evolving to the fixed point; while in loop theory, there exists a new fixed point (saddle point) at $(\phi \simeq 2 / 3,\dot{\phi} = 0)$ in contracting state. We find the two critical solutions starting from the saddle point control the evolution of the solutions starting from the fixed point $(\phi = 1, \dot{\phi} = 0)$ to bounce at small values of scalar field in $0 <\phi < 1$. Other solutions, including the large field inflation solutions, all have the history with $\phi < 0$ which we think of as a problem of the effective theory of loop quantum $f(R)$ theory. Another different thing from loop quantum cosmology with Einstein-Hilbert action is that there exist many solutions do not have bouncing behavior.
\end{abstract}
{\maketitle}

\section{Introduction}

Since the development of observational cosmology and astrophysics in recent years,
such as Cosmic Microwave Background(CMB) \cite{Planck2018a,Planck2018b} and Supernovae surveys\cite{Supernovae1998}, some modified theories of gravity, such as Brans-Dicke theory\cite{Brans1961}, brane-world\cite{Maartens2004}, Einstein-Aether theory\cite{Jacobson2001},etc, have be studied in
variant ways in the literature. The $f (R)$ theory\cite{Sotitiou2010} is a relatively
simple theory among these modified theories, and it has been shown that Starobinsky model\cite{Starobinsky1980}, a special model of $f(R)$ theory, is a candidate inflation model to fit the observations of CMB\cite{Planck2018b}.

Although $f(R)$ theories may give a unified description of early-time inflation with late-time cosmic acceleration, they also inherit the big-bang singularity as in classical general relativity, which means the dyanmics of these modified theories of gravity are not complete.
One the other hand, it is widely believed that the quantum effect of gravity can avoid the big-bang
singularity, and loop quantum gravity (LQG)\cite{Ashtekar2004,Han2007} is such a promising theory of quantum gravity which is non-perturbative and background independent. It has been shown that loop quantum cosmology (LQC)\cite{Ashtekar2006,Bojowald2008,Ashtekar2011a}, the cosmological application of LQG, replaces the big-bang singularity by a big-bounce thus solves the initial
singularity of general relativity. So trying to insert loop quantum correction into $f(R)$ cosmology is an interesting work.

Loop quantum gravity is based on connection-triad variables, and since $f(R)$ theory is a higher-derivative theory, one needs to introduce other variables besides the connection-triad variables. It has been shown that\cite{Sotitiou2010} $f(R)$ theory can be transformed into a special theory of Brans-Dicke theory with a potential for the effective scalar field where the scalar field and its conjugate momentum could be viewed as the other variables. The connection-triad formulation of Brans-Dicke theory has been derived, and the full theory of loop quantum $f(R)$ theory also has been constructed in the work of Zhang and his collaborations\cite{Zhang2011a,Zhang2011b}.
In this paper, we want to study the cosmological part of loop quantum $f(R)$ gravity,
and for explicity, we will consider the Starobinsky model, i.e., the $R^2$-model
with $f (R) = R + \frac{R^2}{6 \mu^2}$ in detail.

There exists a conformal transformation can cast Brans-Dicke theory into an
theory of a scalar field  minimally coupled with geometry, i.e., Einstein theory
with a scalar field, and thus the theory can be loop quantized directly.
We call the theory quantized in this frame the Einstein frame theory, and call the loop quantum theory without conformal transformation the Jordan frame theory. Works on loop quantum $f(R)$ cosmology based on Einstein frame\cite{Amoros2014,Haro2014} or on Jordan frame\cite{Artymowski2013,Zhang2013,Jin2018} both can be seen
in the literature. But in the viewpoint of loop quantum theory, geometry is
quantized and whether making a conformal transformation before quantization or
not would give different quantum theories. In the current work we use Jordan frame
since Jordan frame is closer to the original $f(R)$ theory.
Although some works on Jordan frame have been given in the literature\cite{Artymowski2013,Zhang2013,Jin2018}, we find the dynamics of
cosmos before bouncing is not studied in detail especially there exists a sign
problem of the variable $\cos b$ that has not been addressed.

The structure of the paper is as follows: In Sec \ref{classical}, we first
gives some general aspects of classical $f(R)$ cosmology, and for
compariasion with loop theory, we also consider the dyanmics of $R^2$ theory
in detail. In Sec\ref{loop}, we give the effective equations of loop $f(R)$
theory, and analyse the dynamics of $R^2$ theory by the dynamical system method
and numerical method. In Sec\ref{conclusion}, some discussions and conclusions are given.

\section{Classical theory of $f (R)$ cosmology}\label{classical}
In this section, we first present some results of connection-triad form of $f(R)$ gravity,
and the cosmological dynamics of $f(R)$ cosmology in classical theory, then
focus on an important example, $R^2$ model in detail.

Note that in this paper we take $c = 1 = 8 \pi G$.

\subsection{general considerations}
To get the connection-triad form of $f (R)$ theory whose action is
$S[g]=\frac{1}{2} \int d^4x \sqrt{-g} f(R)$, one can transform the theory
into a special theory of Brans-Dicke theory with zero $\omega$ and the
action of $f(R)$ theory is transformed into
\begin{equation}
  S[g, \phi] = \int d^4 x \sqrt{- g} \left[ \frac{1}{2} \phi R - V (\phi)
  \right],
\end{equation}
where the potential of the scalar field $\phi$ is defined as $V (\phi) = \frac{1}{2} (\phi s - f
(s))$ and $s$ is the solution of $\phi = f' (s)$ (assuming $f'' (s) \neq 0$).
In this formulation, the authors of Ref.\cite{Zhang2011a,Zhang2011b} got the connection-triad form of $f (R)$ theory, where the basic variables are $(A^i_a, E^b_j)$ and $(\phi, \pi)$ and
the Hamilton's constraint density is (other constriants are not given here since
their triviality in cosmology):
\begin{eqnarray}
  \mathcal{H} & = & \frac{\phi}{2} \left[ F^j_{a b} - \left( \gamma^2 +
  \frac{1}{\phi^2} \right) \epsilon_{j m n}  \tilde{K}^m_a \tilde{K}^n_b
  \right] \frac{\epsilon_{j k l} E^a_k E^b_l}{\sqrt{h}} \nonumber\\
  & + & \frac{1}{3} \frac{(\tilde{K}^i_a E^a_i)^2}{\sqrt{h} \phi} + \frac{2
  (\tilde{K}^i_a E^a_i) \pi}{3 \sqrt{h}} + \text{$\frac{\pi^2 \phi}{3
  \sqrt{h}} + \sqrt{h} V$} \nonumber\\
  & + & \sqrt{h} D_a D^a \phi \label{Hamiltion-density}, 
\end{eqnarray}
where $\gamma \tilde{K}^i_a := A^i_a - \Gamma^i_a$.

In the spatially flat isotropic model of cosmology, as in the literature, one first introduces
an elementary cell $\mathcal{V}$ and restricts all integrations to this cell.
Then by fixing a fiducial flat metric $^o q_{a b}$ and denoting the volume the
elementary cell $\mathcal{V}$ by $V_o$, the basic variables of gravity are of
the forms
\begin{equation}
  A^i_a = c V_o^{- 1 / 3 o} \omega^i_a \text{ and } E^a_i = p V_o^{- 2 / 3 o}
  e^a_i,
\end{equation}
and the variables of scalar field are
\begin{equation}
  \phi = \phi \text{ and } \pi = \pi_{\phi} V_o^{- 1} .
\end{equation}
where $(^o e^a_i,^o \omega^i_a)$ are a set of orthogonal triads and co-trials
compatible with $^o q_{a b}$. The Possion brackets become
\begin{equation} \label{Possion-bracket-pre}
\{ c, p \}  =  \frac{\gamma}{3} \text{ and } \{ \phi, \pi \}  =  1.
\end{equation}
In cosmological part, the first three terms in the constraint density
(\ref{Hamiltion-density}) are canceled by each other, and the total Hamiltonian constraint
becomes
\begin{equation}
  C_{\text{cl}} = \frac{2}{\gamma} \frac{c \text{sgn} (p) \pi_{\phi}}{\sqrt{|
  p |}} + \frac{\pi_{\phi}^2 \phi}{3 | p |^{3 / 2}} + | p |^{3 / 2} V (\phi)
  \label{constraint-classical} .
\end{equation}
By the canonical equations and the constraint equation, one can get the
Friedmann equation in $f(R)$ cosmology,
\begin{equation}
  \left( H + \frac{1}{2} \frac{\dot{\phi}}{\phi} \right)^2 = \frac{1}{3}
  \left[ \frac{3}{4} \left( \frac{\dot{\phi}}{\phi} \right)^2 + \frac{V
  (\phi)}{\phi} \right] =: \frac{\phi \rho}{3}
  \label{Friedmann-eqn-cl},
\end{equation}
where $H := \frac{\dot{p}}{2 p}$ is the Hubble constant.

The dynamical equation of scalar field also can be derived as
\begin{equation}
  \ddot{\phi} + 3 H \dot{\phi} + \frac{2}{3} \phi V' (\phi) - \frac{4}{3} V
  (\phi) = 0 \label{scaler-field-eqn-cl} .
\end{equation}
Note that we will restrict the scalar field's region to $\phi>0$ since from the full theory, $\phi$ appears in the denominator of the Hamilton's constraint (\ref{Hamiltion-density}), and
thus in above Eq.(\ref{Friedmann-eqn-cl}), we introduced an ``energy density'':
\begin{equation}\label{rho}
\rho := \frac{1}{\phi} \left[ \frac{3}{4} \left(
      \frac{\dot{\phi}}{\phi} \right)^2 + \frac{V (\phi)}{\phi} \right],
\end{equation}
which is inspired by the theory in the Einstein frame and the evolution
of $\rho$ is
\begin{equation}\label{eqn-of-rho-cl}
\frac{d \rho}{d t} = - \frac{9 \dot{\phi}^2}{2 \phi^3} \left( H +
    \frac{1}{2} \frac{\dot{\phi}}{\phi} \right).
\end{equation}
From Eq.(\ref{Friedmann-eqn-cl}) and since $V(\phi)$ is generally positive, we will find this expression is generally sign fixed which will help analyzing the dynamics of the system.

\subsection{$R^2$ model}

In this paper we focus on the model with $f(R) = R + \frac{R^2}{6 \mu^2}$, called Starobinsky theory or $R^2$-model, so in this subsection, we first consider the classical
dyanamics of this model.

It is easy to find the potential $V (\phi)$ of the scalar
field in Brans-Dicke theory of $R^2$ model is
\begin{equation}\label{V-of-phi}
  V (\phi) = \frac{3}{4} \mu^2 (\phi - 1)^2 ,
\end{equation}
and then the Friedmann's equation and the dynamical equation of scalar field are
\begin{eqnarray}
  \left( H + \frac{\dot{\phi}}{2 \phi} \right)^2 & = & \frac{1}{4} \left[
  \left( \frac{\dot{\phi}}{\phi} \right)^2 + \mu^2 \frac{(\phi - 1)^2}{\phi}
  \right], \\
  \ddot{\phi} + 3 H \dot{\phi} + \mu^2 (\phi - 1) & = & 0.\label{scalar-eqn-cl}
\end{eqnarray}
The right hand of Friedmann equation is positive except the
point $(\phi = 1, \dot{\phi} = 0)$, which is a fixed point from the equation of scalar field(\ref{scalar-eqn-cl}).
This means for a solution which is not a fixed point, it must have $H=H_{+}$ forever or $H=H_{-}$ forever, where
\begin{equation}
  H_{\pm} := - \frac{1}{2} \frac{\dot{\phi}}{\phi} \pm \frac{1}{2}
  \sqrt{\left( \frac{\dot{\phi}}{\phi} \right)^2 + \mu^2 \frac{(\phi -
  1)^2}{\phi}} .
\end{equation}
Moreover, one can find (i) the solution space of the system is time-inverse invariant: if $\phi (t)$ is
a solution, then $\tilde{\phi} (t) := \phi (- t)$ is also a solution; (ii)
and if $\phi(t)$ has $H = H_+\geqslant 0$, then its time-inversion $\tilde{\phi}(t)$ has $H = H_-\leqslant 0$. So in the following, we only consider the solutions having the property $H = H_+$ , and other solutions can be derived by time inversion.

To analyse the dynamics, we rescale the time variable and introduce new variables:
\begin{equation}
  \left\{\begin{array}{lll}
    \tau & := & \mu t,\\
    x & := & 1 - \frac{1}{\phi} \in (-\infty,1),\\
    y & := & \frac{1}{\mu} \frac{\dot{\phi}}{\phi^{3 / 2}}.
  \end{array}\right.
\end{equation}
The dynamicial equations of ${x,y}$ are
\begin{equation} \label{eqns-cl}
  \left\{\begin{array}{lll}
    x' & = & y \sqrt{1 - x},\\
    y' & = & - \frac{3}{2} \sqrt{x^2 + y^2} \frac{y}{\sqrt{1 -
    x}} - x \sqrt{1 - x},
  \end{array}\right.
\end{equation}
where the prime $'$ is for $\frac{d}{d\tau}$. It is interesting to note that these equations are independent on the model parameter $\mu$, thus we know the dynamics of all $R^2$ models with different $\mu$s are similar. Note that we will find loop theory does not have this property.

\begin{figure}[!ht]
  \includegraphics[clip,width=0.48\textwidth,keepaspectratio]{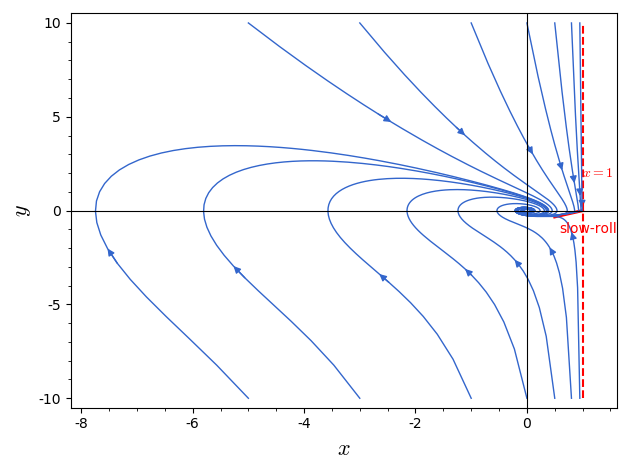}
  \caption{\label{fig-classical}(color online) Phase potrait with $H=H_+$ for Starobinsky
  model, where $x \equiv 1 - \frac{1}{\phi}, y = \frac{1}{\mu}
  \frac{\dot{\phi}}{\phi^{3 / 2}}$.}
\end{figure}

The energy density becomes  $\rho = \frac{3}{4} \mu^2 (x^2 + y^2)$, and its evolution is
\begin{equation}\label{eqn-of-rho-cl-model}
  \frac{d\rho}{d \tau} = - \frac{9}{4} \mu^2 y^2
  \sqrt{\frac{x^2 + y^2}{1 - x}} \leqslant 0.
\end{equation}
From the dynamical equations (\ref{eqns-cl}), one can find the fixed point $(\phi = 1, \dot{\phi} = 0)$, i.e., $(x,y) = (0, 0)$, is the unique fixed point. And since $\rho$ decreases as time flows and the minimum of $\rho$ is taken at the unique fixed point, we know $(x,y) = (0,0)$ is an asympotic stable fixed point and all solutions will evolve to it. To get more information of the system, one needs to solve the dynamical equations (\ref{eqns-cl}), and the results of numerical solutions are shown in the Fig.\ref{fig-classical}. In the figure, we also plot the slow-roll solution, which can be derived as follows. In the case of large field limit, slow-roll approximations require (i) potential dominates energy ( $|y|\ll x$) and (ii) $\dot{\phi}$ varies slow making $y'$ small compared to the other two terms in the second equation of Eq.(\ref{eqns-cl}). The slow-roll approximations transform the dynamics equations(\ref{eqns-cl}) into the slow-roll equation $y=-\frac{2}{3}(1-x)$.
Recall that if one wants to get the solutions with $H = H_- \leqslant 0$, just takes the transformation: $x_- (\tau) := x_+ (- \tau), y_- (\tau) := - y_+(- \tau)$.

\section{Effective theory of loop $f (R)$ cosmology}\label{loop}

\subsection{general considerations}

In this section, we consider the LQC correction to $f (R)$ cosmology. In
loop quantum theory, one uses holonomies of edges and the fluxes on some 2-dim
surfaces, instead of connections and triads. The loop quantum construction of
$f (R)$ cosmology or Brans-Dicke cosmology has been derived in Ref.\cite{Artymowski2013,Zhang2013}, and in this paper we consider the effective theory with holonomy corrections.
Effectively, the holonomy correction is equivalent to do the following
replacement in the constraint (\ref{constraint-classical}):
\begin{equation}
  c \rightarrow \frac{\sin (\bar{\mu} c)}{\bar{\mu}},
\end{equation}
where $\bar{\mu} = \sqrt{\frac{\Delta}{| p |}}$. Thus we get the constraint of
loop quantum cosmology of $f (R)$ cosmology:
\begin{equation}\label{constraint-effective}
C_{\text{eff}} = \frac{2}{\gamma} \frac{\sin (\bar{\mu} c)}{\bar{\mu}}
   \frac{\pi_{\phi}}{\sqrt{| p |}} + \frac{\pi_{\phi}^2 \phi}{3 | p |^{3 / 2}}
   + | p |^{3 / 2} V (\phi)  .
\end{equation}
As in the literature, it is convenient to introduce $\rho_c = \frac{3}{\gamma^2 \Delta} \simeq
0.41 \rho_P$, $b := \bar{\mu} c$ and $v := p^{3 / 2}$. Then the
Possion bracket of $b$ and $v$ is $\{ b, v \} = \frac{1}{2} \sqrt{\frac{3}{\rho_c}}$
from Eq.(\ref{Possion-bracket-pre}), and the constraint $C_{\text{eff}}$ in Eq.(\ref{constraint-effective}) becomes
\begin{equation}
  C_{\text{eff}} = 2 \sqrt{\frac{\rho_c}{3}} \sin (b) \pi_{\phi} + \frac{\phi
  \pi_{\phi}^2}{3 v} + v V (\phi) .
\end{equation}
By the canonical equations and some direct calculations, one can simplify the
constraint equation as
\begin{equation}
  \sin^2 b = \frac{\frac{3}{4}  \dot{\phi}^2 + \phi V (\phi)}{\rho_c} = :
  \frac{\rho_{\text{eff}}}{\rho_c}\label{eqn-of-square-of-sine-b},
\end{equation}
where we introduced the energy density:
\begin{equation}\label{rho-eff}
  \rho_{\text{eff}} := \frac{3}{4} \dot{\phi}^2 + \phi V (\phi),
\end{equation}
which is different from the classical theory's energy density $\rho$, and the relation of the two is $\rho_{\text{eff}} = \phi^3 \rho$. From the
Eq.(\ref{eqn-of-square-of-sine-b}), one can get $\rho_{\text{eff}} \leqslant
\rho_c$, and one may think $\rho_{\text{eff}} = \rho_c$ is just the bouncing
point like in LQC with ordinary scalar field, but we will find it is a necessary condition
but not a sufficient condition.

The Hubble variable $H = \frac{\dot{v}}{3 v}$ in the effective theory satisfies
\begin{equation}\label{Hubble-effective}
  H = \frac{\cos b}{\phi} \left( \sqrt{\frac{\rho_c}{3}} \sin b -
  \frac{\dot{\phi}}{2} \right).
\end{equation}
To compare the Hubble variable's equation(\ref{Hubble-effective}) with the Friedmann equation(\ref{Friedmann-eqn-cl}) in classical theory, one can get
\begin{equation}\label{Friedmann-eqn-loop}
  \left( H + \frac{\cos b}{2} \frac{\dot{\phi}}{\phi} \right)^2 = \frac{\phi
  \rho}{3} \left( 1 - \frac{\rho_{\text{eff}}}{\rho_c} \right).
\end{equation}
Since $\cos b = \pm \sqrt{1 - \frac{\rho_{\text{eff}}}{\rho_c}}$, we can find when
$\rho_{\text{eff}} \ll \rho_c $ and $\cos b > 0$, the Friedmann equation of effective theory closes to classical theory(cf.Eq.(\ref{Friedmann-eqn-cl})), but those cases with $\cos b<0$ do not
close classical theory. The equation of scalar field is also modified
\begin{equation}\label{scalar-eqn-effective}
  \ddot{\phi} + 3 H \dot{\phi} + \frac{2}{3} \phi V' + \frac{2}{3} V (1 - 3
  \cos b) = 0.
\end{equation}
From the Friedmann equation and the equation of scalar field, we have known loop
quantum $f(R)$ cosmology needs more conditions to get classical theory than the
LQC with Einstein-Hilbert action, where the energy condition $\rho\ll\rho_c$ is enough to get classical theory.

Since the potential function of scalar field is not given explicitly, the
dynamics of the theory can not be analysed in detail, and we consider the $R^2$ model
in the next subsection.

\subsection{$R^2$ model}

The potential of scalar field in $R^2$ model is given by Eq.(\ref{V-of-phi}),
and we also make some rescalings:
\begin{equation}
  \tau = \mu t \text{ and } y = \frac{\dot{\phi}}{\mu} \equiv \frac{d \phi}{d
  \tau}\equiv \phi'.
\end{equation}
Then the constraint equation (\ref{eqn-of-square-of-sine-b}) becomes
\begin{equation}\label{eqn-of-square-sine-b-model}
  \sin^2 b = \frac{3 \mu^2}{4 \rho_c} [y^2 + \phi (\phi - 1)^2] =:
  \epsilon \tilde{\rho} (\phi, y) ,
\end{equation}
where $\tilde{\rho} := y^2 + \phi (\phi - 1)^2$, and $\epsilon = \frac{3 \mu^2}{4 \rho_c} \sim 10^{- 12}\ll 1$ is chosen to fit with the observations\cite{Planck2018b}. The Hubble variable satisfies
\begin{equation}
  H = \frac{\mu}{2 \sqrt{\epsilon}} \frac{\cos b}{\phi} \left( \sin b -
  \sqrt{\epsilon} y \right) \label{Hubble-effective-model},
\end{equation}
and the dynamical equations of scalar fields are
\begin{equation}\label{eqn-phi}
  \frac{d \phi}{d \tau} = y,
\end{equation}
\begin{equation}\label{eqn-y}
  \frac{d y}{d \tau} = - 3 \frac{H}{\mu} y - (\phi - 1) \left[ \phi +
  \frac{\phi - 1}{2} (1 - 3 \cos b) \right] .
\end{equation}

To solve the system, we view $(\phi, y)$ as the dynamical variables,
and $b$ is not viewed as a dynamical variable since
the Eq.(\ref{eqn-of-square-sine-b-model}) almost determines $b$ except
the sign of $\sin b$ and $\cos b$.
And for $\sin b$, we find the sign of $\sin b$ never changes since
$\tilde{\rho} \geqslant 0$ where '$=$' only happens at the fixed point
$(\phi = 1, y = 0)$(cf. Eq.(\ref{eqn-phi}),(\ref{eqn-y})).
Moreover, the same with classical theory, the space of
solutions also has the property of time-inverse symmetry:
$(\phi (\tau), y (\tau), b (\tau)) \rightarrow (\phi (- \tau), - y (- \tau), -
b (- \tau))$. Thus we can safely only consider the subset of the solution
space where every solution (except the fixed point) has
\begin{equation}
  \sin b = \sqrt{\epsilon \tilde{\rho}} > 0,
\end{equation}
which makes $b \in (0, \pi)$. After this, we can split the range of $b$ into $\left(
0, \frac{\pi}{2} \right)$ and $\left( \frac{\pi}{2}, \pi \right)$, then
consider each part separately thus the sign of $\cos b = \pm
\sqrt{1 - \epsilon \tilde{\rho}}$ is also solved.

In the following, we first consider the behavior near $b = \frac{\pi}{2}$ since
it is related to bouncing states, then we try to find all fixed points and
analyze the dynamics near those fixed points. The singular edge $\phi = 0$
is important to the system, and we will analyse the dynamics near the singular edge
$\phi = 0$. After these analytical analyses, we confer to the numerical method.
It may be noted that the analytical analyses will help to the numerical analyse
since the complexity of the system.

\subsubsection{The behavior near $b = \frac{\pi}{2}$}\label{bouncing-point}

The range of $b$ is $(0, \pi)$, and by the equation of Hubble variable
(\ref{Hubble-effective-model}), it would have $H = 0$ when $b =
\frac{\pi}{2}$. But not all the points with $b = \frac{\pi}{2}$
are the bouncing points. A bouncing point is a point where the collapsing world
is transforming into expanding world, so the Hubble variable $H$ should
evolve from negative to positive at the point. Since $\sin b - \sqrt{\epsilon} y =\sqrt{\epsilon} (\sqrt{\tilde\rho}-y) > 0$
(except the point with $\phi = 1$) and by the equation of Hubble variable(\ref{Hubble-effective-model}),
we have $\text{sign} (H) = \text{sign} (\cos b)$, which means a bouncing
solution has a decreasing $b$ as time flows. Then from the canonical equation of $b$,
\begin{equation}\label{eqn-of-b}
  \frac{d b}{d \tau} = \frac{3 y}{2 \phi} \left( \sin b - \sqrt{\epsilon} y
  \right),
\end{equation}
we know the points with $y < 0$ and $b= \frac{\pi}{2}$ are the bouncing points.
Fig.\ref{Three-classes-solutions-a} gives solutions of this case. But for the solutions
with $y > 0$ at $b = \frac{\pi}{2}$, they evolve from expanding states to collapsing states.
The case(b1) in Fig.\ref{Two-classes-solutions-b} gives such a case.

\subsubsection{The fixed points and the dynamics near them}

The fixed points and the behavior near the fixed points can help to understand the global behavior of a dynamical system. In this part, we consider related problems.

Fixed points satisfy the equations
\begin{equation}
  \left\{\begin{array}{l}
    y = 0,\\
    (\phi - 1) \left[ \phi + \frac{\phi - 1}{2} (1 - 3 \cos b) \right] = 0.
  \end{array}\right.
\end{equation}
There are three solutions of these equations. Two of them
are easy to find, which also appear in classical theory: $(\phi = 1, y = 0)$ with $b = 0$ or $\pi$. There is a new fixed point satisfying $y = 0$ and $\phi + \frac{\phi - 1}{2} (1 - 3 \cos b) = 0$,
i.e., $\left( \phi = \hat{\phi} \simeq \frac{2}{3} -
\frac{\epsilon}{162} + o (\epsilon), y = 0 \right)$ with $b = \arccos\left(\frac{3 \hat{\phi} -
  1}{3 (\hat{\phi} - 1)}\right) \in \left( \frac{\pi}{2}, \pi \right)$.
[To solve $\phi=\hat\phi$, let $x = \phi - 1$ and
substitute the expression of $\cos b = \pm \sqrt{1 - \epsilon \tilde{\rho}}$,
then $x$ satisfies $x = - \frac{1}{3} - \frac{3}{4} \epsilon (x + 1) x^4$ who has a unique solution approximated by $x = - \frac{1}{3} - \frac{\epsilon}{162} + o(\epsilon)$ as $\epsilon \sim 10^{- 12} \ll 1$.] The appearance of new fixed point makes the dynamics more complicate than classical theory.

Next we consider the stability of these fixed points or the behaviour near the fixed points.

(i) 1st fixed point: $(\phi = 1, y = 0)$ with $b = 0$. Inspired from classical
theory, there is a Lyapunov function $\rho :=
\frac{\rho_{\text{eff}}}{\phi^3}$ which is positive except the fixed point. It can be
shown that near this fixed point, $\rho$ decreases as time flows (by Eq.(\ref{eqn-of-square-sine-b-model}),(\ref{eqn-of-b})):
\begin{eqnarray*}
  \frac{d \rho}{d \tau} & = & 3 \rho_c \frac{y}{\phi^4} \sin (b) \left[ \cos b
  \left( - \sqrt{\epsilon} y + \sin b \right) - \sin b \right]\\
  & = & 3 \rho_c \frac{y}{\phi^4} \sin (b) \left[ - \sqrt{\epsilon} y +
  \mathcal{O} (\sin^2 b) \left( \sin b - \sqrt{\epsilon} y \right) \right]\\
  & \simeq & - 3 \rho_c \sqrt{\epsilon} \frac{y^2}{\phi^4} \sin b \leqslant
  0,
\end{eqnarray*}
so this fixed point is asymptotically stable and a sink of the system, which is the same
with classical theory.

(ii) 2nd fixed point: $(\phi = 1, y = 0)$ with $b = \pi$. The function $\rho =
\frac{\rho_{\text{eff}}}{\phi^3}$ is not a Lyapunov function since the dynamics of
loop theory is quite different from classical theory when $\cos b < 0$.
Thus we need to consider the dynamical equations near the fixed point. Let $x = \phi
- 1$ and introduce the polar coordinates $x = r \cos \theta, y = r \sin
\theta$, then the equations of $r,\theta$ are approximated as
\[ \left\{\begin{array}{lll}
     \frac{d \theta}{d \tau} & = & - 1 + \mathcal{O} (r)\\
     - \frac{d r}{d \theta} & = & \frac{3}{2} r^2 (\sin^3 \theta + \sin^2
     \theta - 2 \sin \theta) + \mathcal{O} (r^3)
   \end{array}\right., \]
The first equation gives that $\theta$ decreases as time flows.
Then by the second equation, one can find after a period of $\theta$, i.e., $\theta$ evolves from $\theta_0$ to $\theta_0 - 2 \pi$, $r$ would increase,
\begin{eqnarray*}
  &  & r (\theta_0 - 2 \pi) - r (\theta_0) = \int_{\theta_0}^{\theta_0 - 2
  \pi} \frac{d r}{d \theta} d \theta\\
  & \simeq & \frac{3}{2} r (\theta_0)^2 \int_{\theta_0 - 2 \pi}^{\theta_0}
  [\sin^3 \theta + \sin^2 \theta - 2 \sin \theta] d \theta\\
  & = & \frac{3 \pi}{2} r (\theta_0)^2 > 0.
\end{eqnarray*}
This proves that the fixed point is asymptotically unstable and a source of the system. Note that
we also can use this method to analyze the stability of the 1st fixed point which gives the same result as above.

(iii) 3rd fixed point: $\left( \phi = \hat{\phi} \simeq \frac{2}{3} -
\frac{\epsilon}{162} + o (\epsilon), y = 0 \right)$ with $b = \arccos\left(\frac{3 \hat{\phi} -
1}{3 (\hat{\phi} - 1)}\right) \in \left( \frac{\pi}{2}, \pi \right)$.
Let $\xi = \phi - \hat{\phi},\eta = y$, and the dynamics near the fixed point to linear order is
\[ \left\{\begin{array}{lll}
     \frac{d \xi}{d \tau} & = & \eta\\
     \frac{d \eta}{d \tau} & = & \xi + \frac{\sqrt{6}}{4} \eta
   \end{array}\right., \]
whose general solution is
\[ \left[ \begin{array}{l}
     \xi\\
     \eta
   \end{array} \right] = c_+ g_+ (\tau) + c_- g_- (\tau), \]
where
$g_{\pm} (\tau) := e^{\tau \lambda_{\pm}} \left[ \begin{array}{l}
  1\\
  \lambda_{\pm}
\end{array} \right],$
$\lambda_{\pm} = \frac{\sqrt{6}}{8} \left( 1 \pm \sqrt{\frac{35}{3}} \right)$
and $c_{\pm} \in \mathbb{R}$. Since $\lambda_+ > 0, \lambda_- < 0$, we know
this fixed point is a saddle point.

Since there are only one source (in collapsing world) and only one sink (in expanding world), one may think if there exists no limit cycle, then all solutions, except the fixed points and the critical solutions $\pm g_{\pm}$ (here and later,
what we say $\pm g_{\pm}$ is to mean the global solutions of $\phi=\xi+\hat\phi,y=\eta$ where $[\xi,\eta]^T\simeq\pm g_{\pm}$ when the points are near the saddle point.), would
start from the source and evolve to the sink, like the dynamics in the LQC of Einstein-Hilbert action \cite{Singh2006}. However, we will find this is not the case.
The problem is the region $\phi > 0$ we restrict to is not dynamically closed, i.e., many solutions are from $\phi < 0$ and cross the singular edge $\phi = 0$ to $\phi>0$ which is shown in the following part.

\subsubsection{The dynamics near the singular edge $\phi = 0$}

In this part, we consider the behavior of the system near the singular edge
$\phi = 0$.
Let $y \neq 0$ and assume $\phi \sim 0_+$, then by the constraint equation
(\ref{eqn-of-square-sine-b-model}), one can get $\sin b \simeq
\sqrt{\epsilon} | y | \left( 1 + \frac{\phi}{2 y^2} \right)$. Substitude this
expression into $H$ in Eq.(\ref{Hubble-effective-model}), one can find the
Hubble variable satisfies
\[ H \simeq \left\{\begin{array}{ll}
     \frac{\mu}{4 y} \cos b, & \text{if } y > 0\\
     \mu y \frac{\cos b}{\phi}, & \text{if } y < 0
   \end{array}\right. . \]
Then the dynamics of $y$ in Eq.(\ref{eqn-y}) becomes
\[ \frac{d y}{d \tau} = \left\{\begin{array}{ll}
     - \frac{1}{2} + \frac{3}{4} \cos b + \mathcal{O} (\phi), & \text{if } y >
     0,\\
     - \frac{3 y^2 \cos b}{\phi} + \mathcal{O} (\phi^0), & \text{if } y < 0.
   \end{array}\right. \]
In the case of $y < 0$, we have $\frac{d y}{d \tau} \rightarrow \pm \infty$ when
$\phi \rightarrow 0_+$, so no solution can cross the edge $\phi
= 0$. However for the solutions with $y > 0$, $\frac{d y}{d \tau}$ is finite when $\phi
\rightarrow 0_+$ which means there exists solutions can cross
the edge $\phi = 0$ from $\phi<0$. We think this is a bad thing and
it is not a special thing of $R^2$ theory but common for the general $f (R)$ theory where $\frac{d \dot{\phi}}{d t} \simeq - \frac{2 V}{3} (1 - 3 \cos b)$ when $\dot{\phi} > 0$ and $\phi \rightarrow 0_+$. Note that $\phi=0$ also can be crossed in classical theory, but from the definition of energy density(\ref{rho}), these points of $\phi=0$ are related to the singularity of the universe with $\rho=\infty$.

Since $\phi = 0$ is not the dynamical edge of the system, when we try to get the total
phase portrait by numerical analyze as in the following subsubsection, we need be careful of the edge of $\phi=0$.

\subsubsection{Numerical analyze}

\begin{figure}[!ht]
  \includegraphics[clip,width=0.48\textwidth,keepaspectratio]{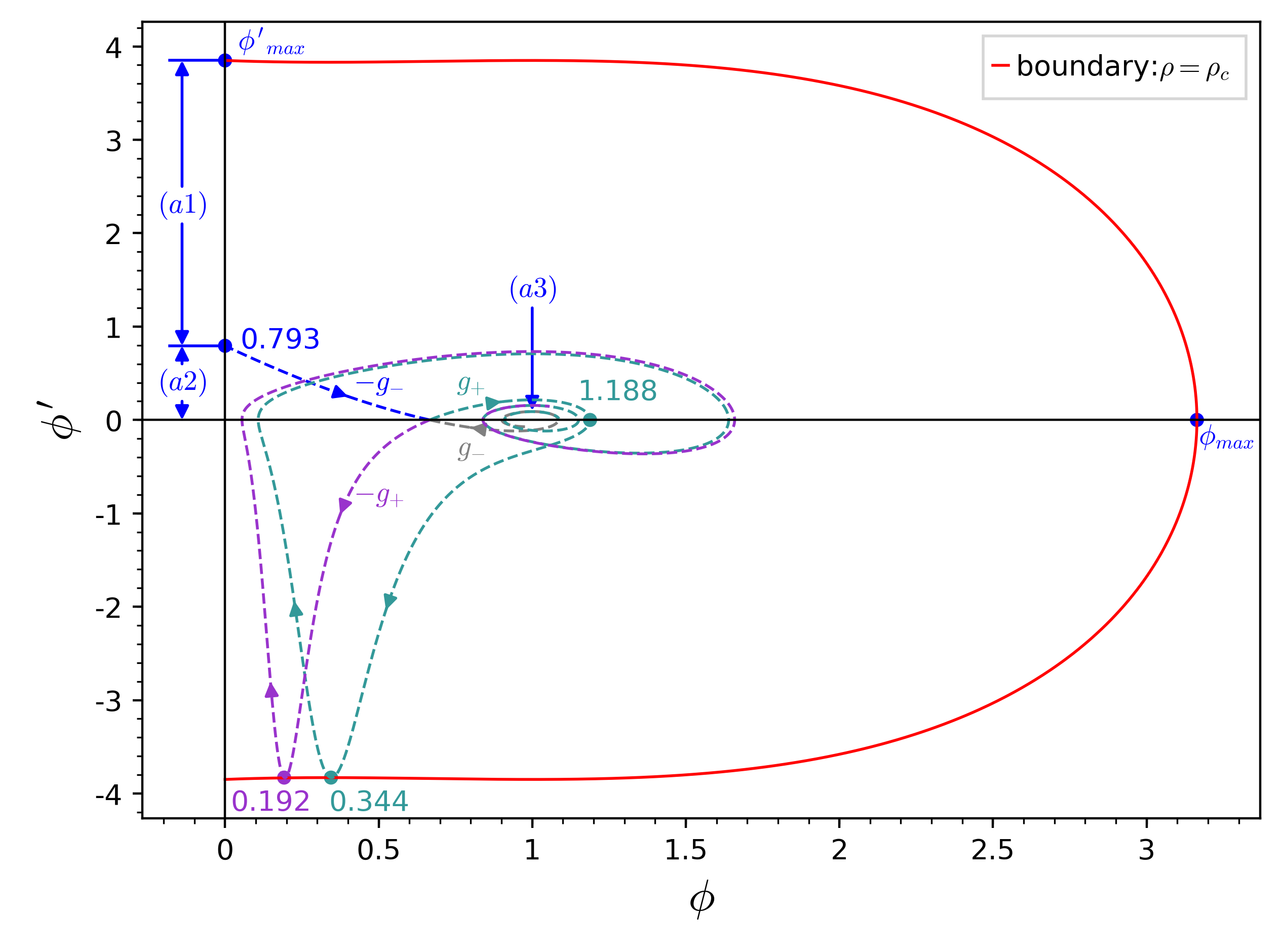}
  \caption{\label{critical-solutions-a} (color online)
    ($\sigma \equiv \frac{\mu}{\sqrt{\rho_c}} = 0.3$)
    The global behavior of critical solutions $\pm g_{\pm}$ from or to the saddle point $(\phi \simeq \frac{2}{3}, \phi'=0)$. The two solutions $\pm g_+$ restrict the solutions starting from source $(1,0)$ can only bouncing at $0<\phi<1$. We also show the initial points (at $\phi=0$) of the three classes of solutions: (a1), (a2) and (a3).}
\end{figure}

The critical solutions $\pm g_{\pm}$ related with the saddle point can grab
the global behavior of the phase portrait, so we first consider the behaviors of
these four solutions, which are given in Fig.\ref{critical-solutions-a}, where for pretty plotting we choose $\sigma = \frac{\mu}{\sqrt{\rho_c}} = 0.3$ and $\epsilon = \frac{3}{4} \sigma^2 = 0.0675
\ll 1$ which does not break the condition when solving the saddle point.
The real model with $\epsilon \sim 10^{- 12}$ will be given in the last of this subsubsection whose results are not quite different.

From Fig.\ref{critical-solutions-a}, we know: (1) the solution $- g_-$ starts at $(\phi = 0, y =
0.793)$ with $\cos b < 0$; (2) $g_-$ starts from the source $(1, 0)$ with $b = \pi$;
(3) $g_+$ evolves over the source $(1,0)$ to $(1.188, 0)$ and bounces at $\phi =
0.344<1$, then evolves to the sink $(1, 0)$; (4) $- g_+$ bounces at $\phi =
0.192$ and also evolves to the sink $(1, 0)$; (5) one important thing is that the two solutions $\pm g_+$ together with the boundary $\rho = \rho_c$ make a closed region where the source $(1, 0)$ sits in which means all solutions starting from the source will bounce at $\phi \in (0.192, 0.344)\subset (0,1)$, which also means all of these solutions are not large field inflationary solutions.

By these four critical solutions and the source point $(1,0)$, we know there are three classes soluitons whose initial values are classified as

(a1) $\phi = 0, \phi' \equiv y > 0.793$ with $\cos b < 0$;

(a2) $\phi = 0, 0 < \phi' < 0.793$ with $\cos b < 0$;

(a3) $\phi \sim 1, \phi' \sim 0$ with $b \sim \pi$.

\begin{figure}[!ht]
  \includegraphics[clip,width=0.48\textwidth,keepaspectratio]{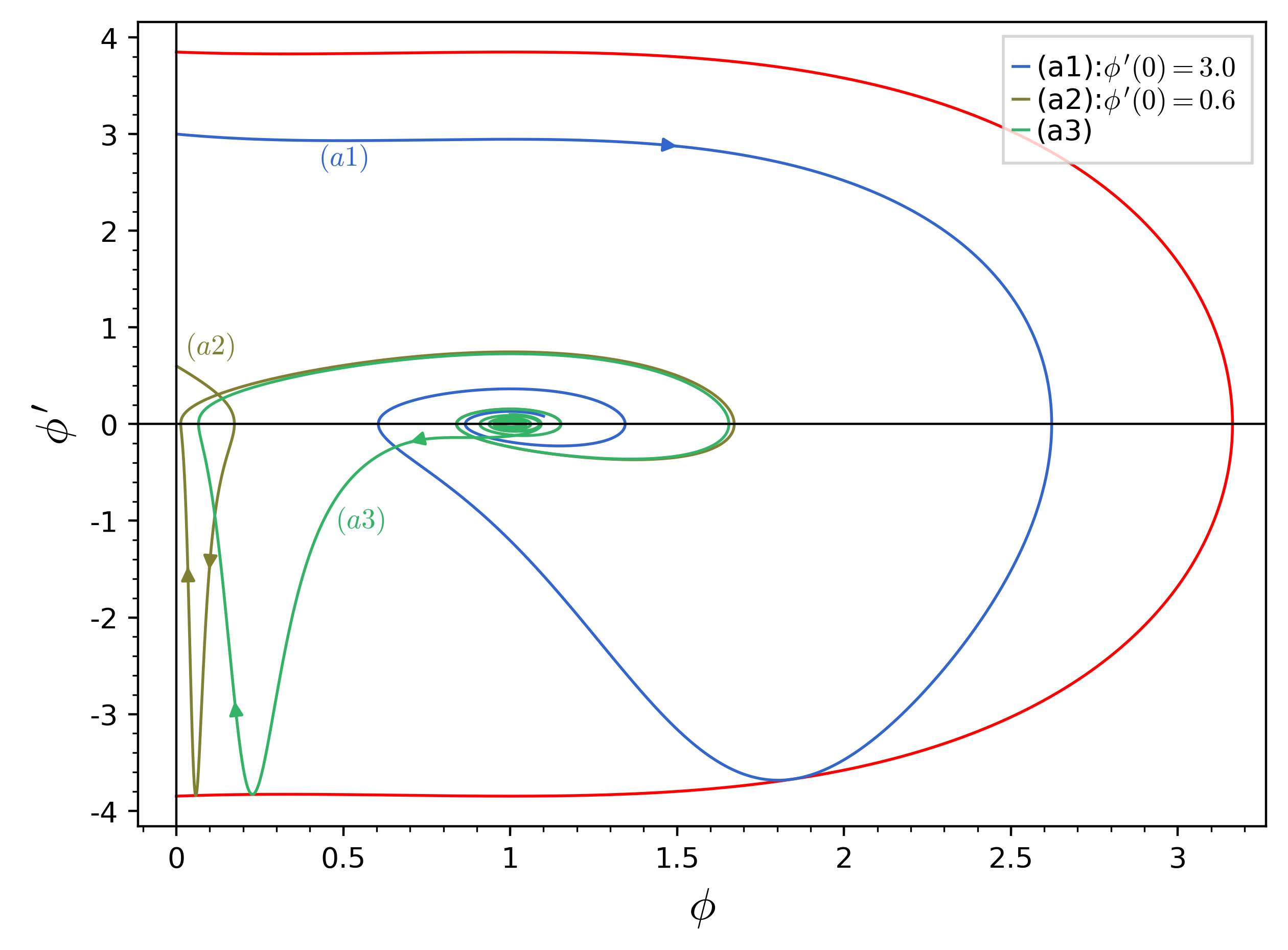}
  \caption{\label{Three-classes-solutions-a}(color online)
    ($\sigma \equiv \frac{\mu}{\sqrt{\rho_c}} = 0.3$)
    Three examples of solutions (a1),(a2) and (a3) whose initial values have $\cos b<0$. The solutions in case (a1) bounce at large field $\phi>0.344$, while the solutions in the cases of (a2) and (a3) all bounce at small field $0<\phi<0.344$. }
\end{figure}

Each a typical solution of these three cases are shown in Fig.\ref{Three-classes-solutions-a}.

\begin{figure}[!ht]
  \includegraphics[clip,width=0.48\textwidth,keepaspectratio]{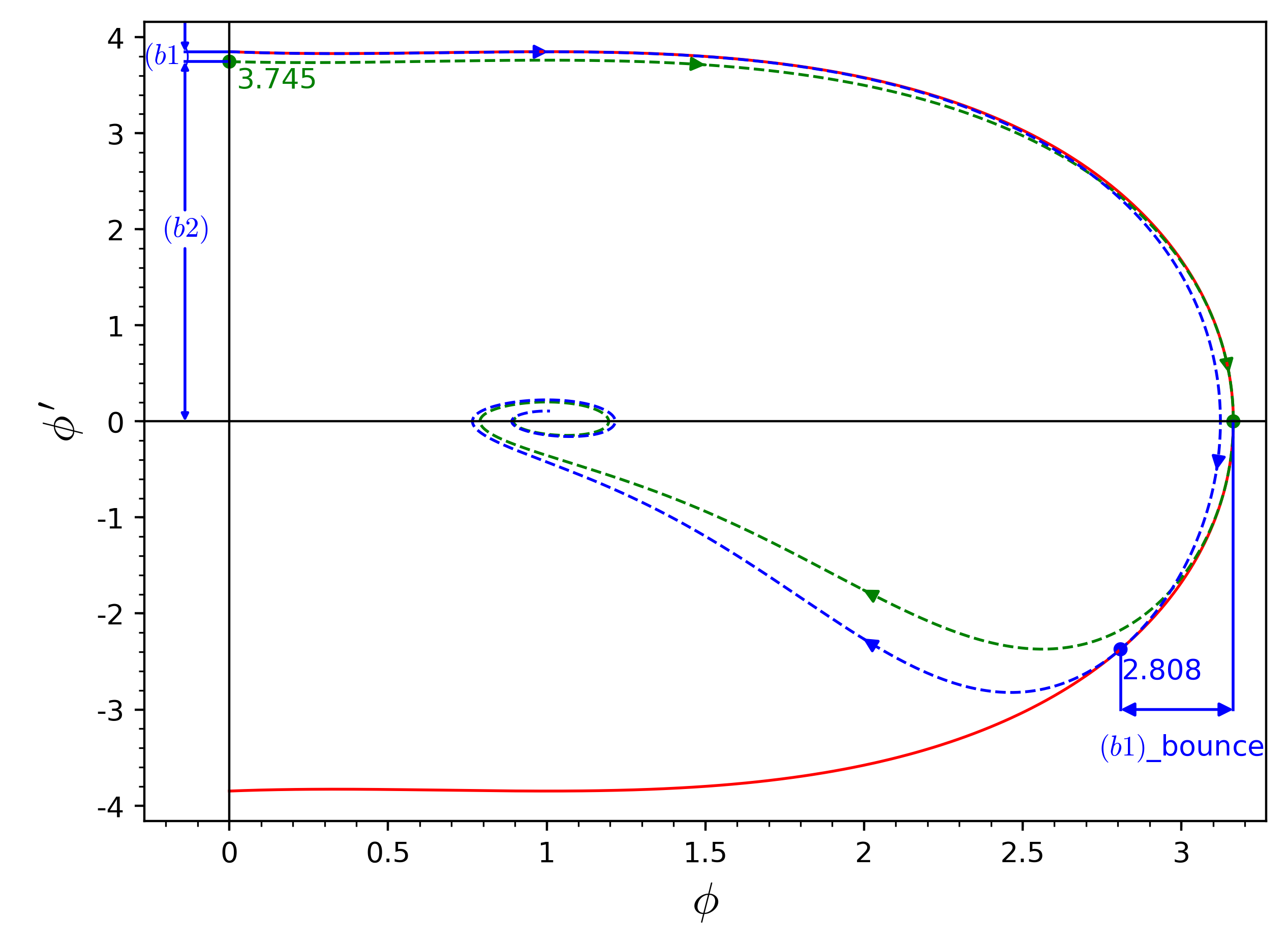}
  \caption{\label{Two-critical-solutions-b}(color online)
    ($\sigma \equiv \frac{\mu}{\sqrt{\rho_c}} = 0.3$)
    Two critical solutions: one starts at $\phi'=\phi'_{max}$ and bounces at $\phi=2.808$, the other one starts at $\phi'=3.745$ and bounces at $\phi=\phi_{max}$. We also show the initial points (at $\phi=0$) of the two classes of solutions: (b1) and (b2).}
\end{figure}

These solutions don't cover all the bouncing points on the boundary $\rho=\rho_c$, since in case
(a1), even we let $\phi'(0)$ be the maximum $\phi'_{\max}$, it only bounces at $\phi \simeq 2.808<\phi_{max}$ (see Fig.\ref{Two-critical-solutions-b}). We find for those solutions bouncing at $\phi \in (2.808, \phi_{\max})$, the initial values would have $\cos b > 0$ at edge $\phi =0$. But these solutions are also not all cases and only cover initial values $\phi'> 3.745$ with $\phi =0$, and the solution of initial value $(\phi = 0, \phi' = 3.745)$ with $\cos b > 0$, bounces at $\phi_{max}$. So we find the initial values with $\cos b>0$ are split into two classes(see Fig.\ref{Two-critical-solutions-b}):

(b1) $\phi = 0, \phi'> 3.745$ with $\cos b > 0$;

(b2) $\phi = 0, 0 < \phi' < 3.745$ with $\cos b > 0$.

\begin{figure}[!ht]
  \includegraphics[clip,width=0.48\textwidth,keepaspectratio]{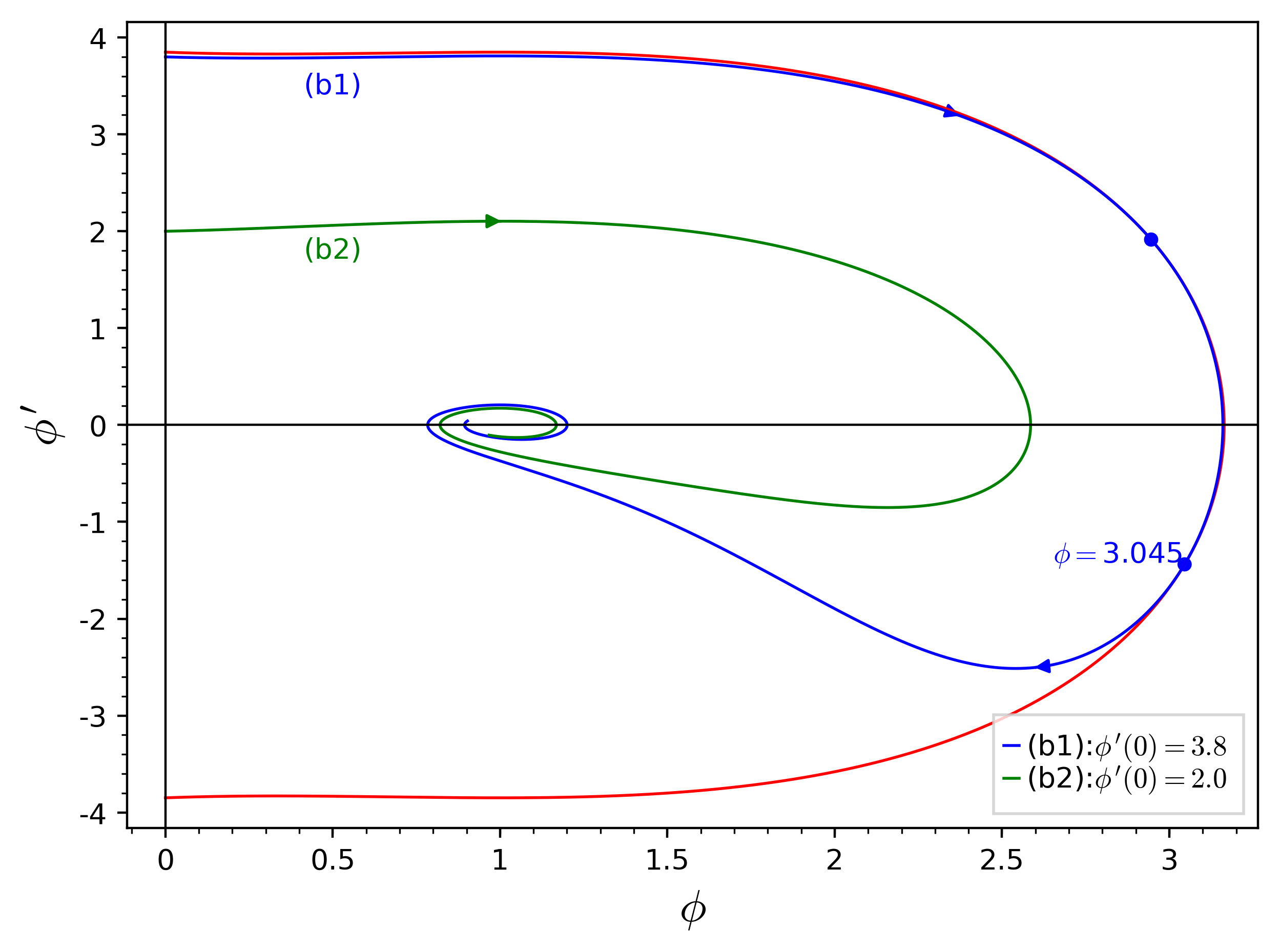}
  \caption{\label{Two-classes-solutions-b}(color online)
    ($\sigma \equiv \frac{\mu}{\sqrt{\rho_c}} = 0.3$)
    Two examples of the solutions of classes (b1) and (b2). The solution of (b1) approaches to $b=\frac{\pi}{2}$ twice (the first one is not bouncing point and the second one is) while the solution of (b2) never bounces.}
\end{figure}

For a solution in case (b1) (cf. Fig.\ref{Two-classes-solutions-b} and Eq.(\ref{Hubble-effective-model})(\ref{eqn-of-b})),
it first expands with $b$ increasing, and when $b = \frac{\pi}{2}$, it begins to collapse but $b$ still increases until $y=0$, and when $b$ decreases to $\frac{\pi}{2}$, it bounces at $\phi \in (2.808,\phi_{\max})$, finally it approaches the sink $(1,0)$ with $b = 0$. On the other hand, for the solutions in case (b2), $b$ also increases initially but never reaches $b=\frac{\pi}{2}$, and when $y=0$, $b$ begins to decrease to zero. These solutions have no bouncing point.

\begin{figure}[!ht]
  \includegraphics[clip,width=0.48\textwidth,keepaspectratio]{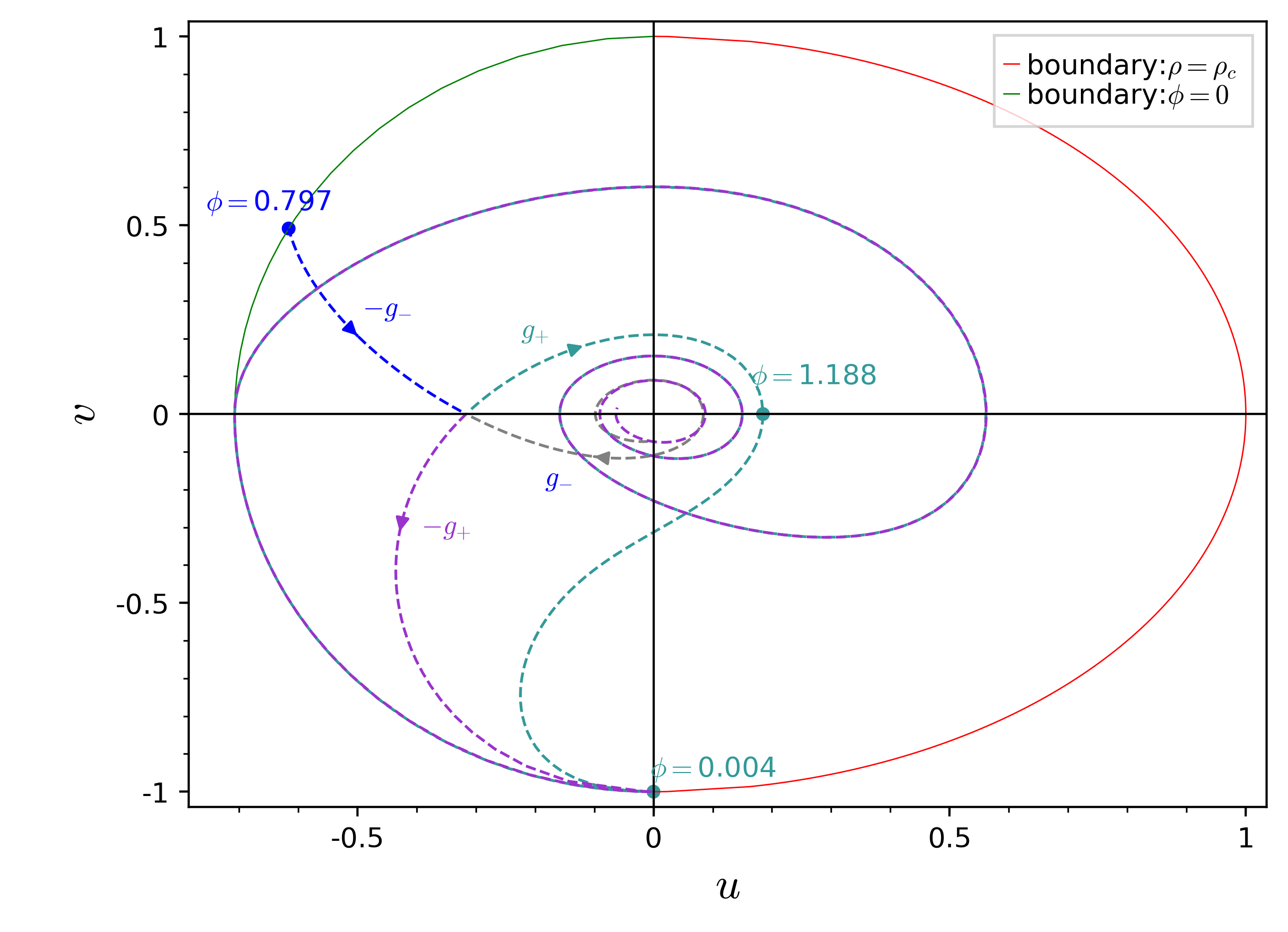}
  \caption{\label{critical-solutions-a-real}(color online)
    ($\sigma \equiv \frac{\mu}{\sqrt{\rho_c}} = 10^{- 6}$)
    Critical solutions in $(u,v)$-space: $\pm g_{\pm}$, which give the classes of solutions (a1), (a2) and (a3). $u,v$ are defined in Eq.(\ref{Poincare-map}).}
\end{figure}

\begin{figure}[!ht]
  \includegraphics[clip,width=0.48\textwidth,keepaspectratio]{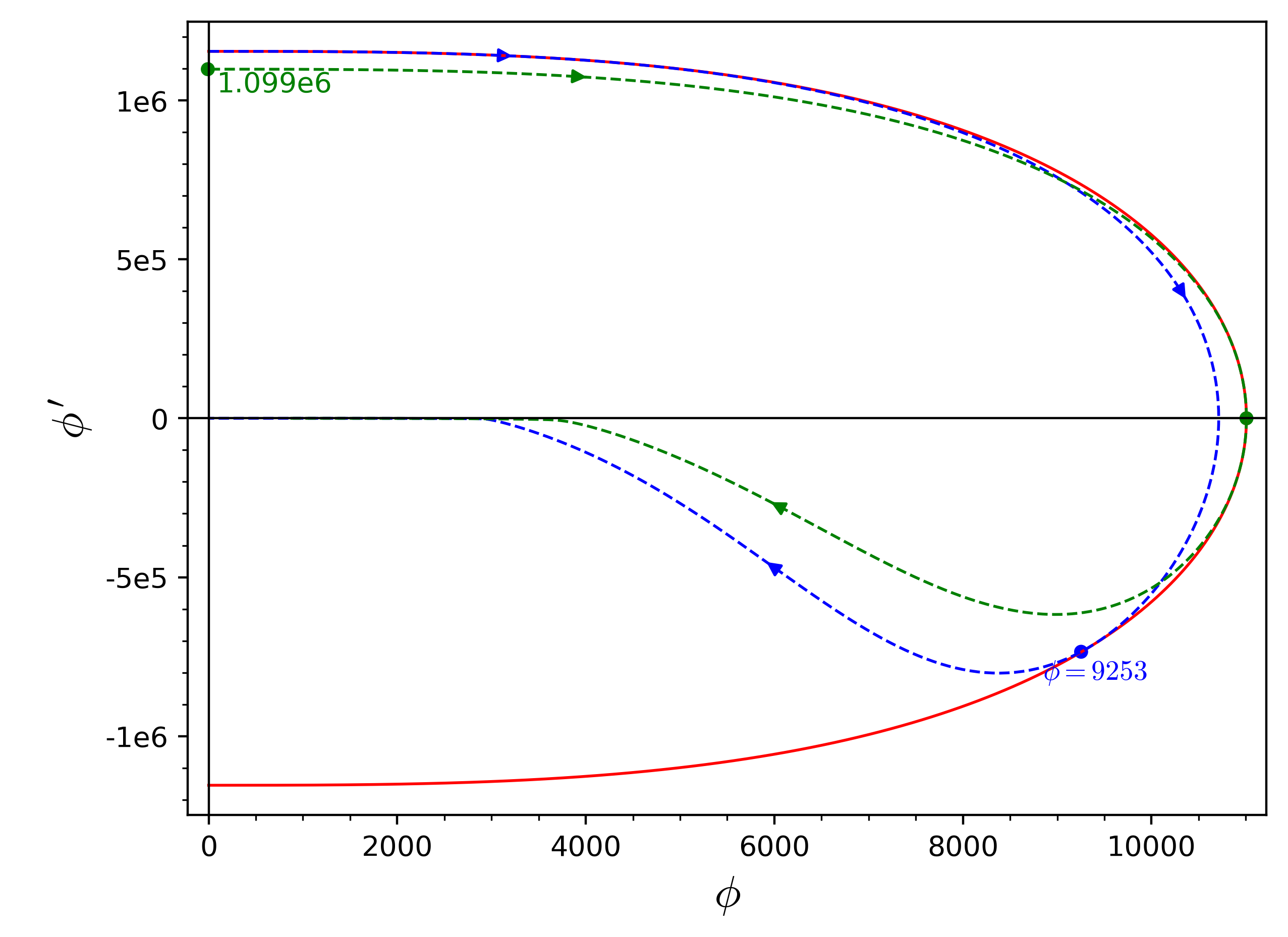}
  \caption{\label{critical-solutions-b-real}(color online)
    ($\sigma \equiv \frac{\mu}{\sqrt{\rho_c}} = 10^{- 6}$) Critical solutions which give the classes of solutions (b1) and (b2).}
\end{figure}

Things are qualitatively same for the real model where $\sigma \equiv
\frac{\mu}{\sqrt{\rho_c}} \sim 10^{- 6}$ with $\epsilon \sim 10^{- 12}$, and
we only gives the behaviors of the critical solutions
in Fig.\ref{critical-solutions-a-real} and Fig.\ref{critical-solutions-b-real}.
In Fig.\ref{critical-solutions-a-real}, for pretty plotting, we use the Poincare's sphere method\cite{Perko2013,Belinsky1985} to map $(x \equiv \phi - 1, y)$ into $(u, v)$:
\begin{equation}\label{Poincare-map}
u := \frac{x}{\sqrt{1 + (x^2 + y^2)}} \text{ and }   v := \frac{y}{\sqrt{1 + (x^2 + y^2)}},
\end{equation}
since the values of $\phi$ and $\phi'$ can be quite large with order of $10^4$ and $10^6$ while the fixed points are of order 1.
Quantitatively, the solution $ g_{+}$ evolves over the source and bounces at $\phi=0.004$ which is much smaller than the previous case, so the solutions from source are also not inflation solutions while other solutions, including inflation solutions, have the history with $\phi < 0$. From the critical solutions in Fig.\ref{critical-solutions-b-real} and the analyze of the case $\sigma=0.3$, we know that for the solutions whose initial values are $\phi=0,\phi'>1.099\times 10^6$ with $b<\frac{\pi}{2}$, they will bounce at $\phi\in(9523,\phi_{max})$, and for the solutions with initial values of $\phi=0,\phi'<1.099\times 10^6$, no bouncing phenomenon happens.

Let's make some comments to end this section: There are five classes of solutions of this system (not including the critical solutions). Only case (a3) has the similar property  with loop theory in Einstein-Hilbert action, i.e., the universe starts from low energy density, bounces at some point, and then goes back to low energy density state. But these solutions do not belong to large field inflation models, since they bounce at low values of the field $\phi$. On the other hand, those large field inflation solutions are in the cases of (a1), (b1) or (b2) where all solutions come from $\phi\leqslant 0$ violating the condition of full theory. Moreover, the solutions in case (b2) even have no bouncing behavior which is not foreseen in ordinary theory of LQC.

\section{Conclusions}\label{conclusion}
The $f(R)$ theory of gravity is a class of modified theories, which can explain some phenomena from cosmology and astrophysics, and has got much attention in the literature.
In the paper, we considered the dynamics of $f(R)$ theory in loop quantum cosmology in Jordan frame. We first considered the classical dynamics of $f (R)$ cosmology, and then considered the effective dynamics of loop quantum cosmology of $f(R)$. We focus on an important model of $f(R)$ theory, i.e., $R^2$ theory. The classical theory is easy to analyze since there exists a global inequality (\ref{eqn-of-rho-cl-model}) of energy density to control the dynamics and only one fixed point in collapsing world or expanding world. However the effective LQC of the theory is some complicate.

Firstly, we find a new fixed point in collapsing world which is a saddle point. Then by analyzing the critical solutions of the saddle point and considering the range of the value of scalar field of bouncing points, we get to know there are five classes of solutions of the system. Only one case of solutions has the property that the universe starting from low energy density states, bouncing at some point and then evolving back to the low energy density states, but these
solutions are not large field inflationary solutions since they bounce at small fields. Other cases of solutions all have a history with $\phi<0$, which may be a problem of the theory since from the full theory of $f(R)$ theory $\phi$ could not be zero, and the potential part of energy density defined in Eq.(\ref{rho-eff}) should be positive. We also note that this phenomenon is not only a special case of $R^2$ model, but also a phenomenon of general $f(R)$ theory. Another quite different thing is that not all solutions have the bouncing behavior, i.e., the solutions in case(b2). This means putting initial values at bouncing points used in the loop theory of Einstein-Hilbert action \cite{Ashtekar2010,Ashtekar2011b,Chen2015} may not be reasonable when considering the e-foldings and the probabilities of happening of inflation in $R^2$ model. And the initial values given at $\phi =0$ may be reasonable since only the case(a3) of solutions, which also are not inflationary solutions, are not the included.

At last, we want to say that the problem of $\phi=0$ may be solved in the general model of Brans-Dicke theory with $\omega>0$ since when considering the dynamics near the edge $\phi\sim 0_+$, one can find the kinetic energy of the energy density would make $\dot\phi\rightarrow\infty$, but the full dynamics of such model should be considered in detail which is not the work of the current work and may be considered in the future work.

\acknowledgments The author thanks Dr. Yu Han for helpful discussions. This work is supported by the Nanhu Scholars Program for Young Scholars of Xinyang Normal University.

\end{document}